\documentclass{article}
\usepackage{spconf,amsmath,graphicx,hyperref}

\usepackage{amssymb,amsfonts,bm}
\usepackage{subfig} 

\usepackage{placeins}

\usepackage[subtle, mathspacing=normal, wordspacing=normal]{savetrees}

\newcommand{\field}[1]{\mathbb{#1}} 

\newcommand{\trace}[1]{\textrm{\textbf{tr}} \left ( {#1} \right )}

\newcommand{\BRAK}[1]{\left [ {#1} \right ]}
\newcommand{\CBRAK}[1]{\left \{ {#1} \right \} }
\newcommand{\ABS}[1]{\left | {#1} \right | }
\newcommand{\PAREN}[1]{\left ( {#1} \right )}

\newcommand{\argmax}[1]{\underset{#1}{\operatorname{arg}\,\operatorname{max}}\;}
\newcommand{\argmin}[1]{\underset{#1}{\operatorname{arg}\,\operatorname{min}}\;}

\newcommand{\NORM}[1]{\left \| {#1} \right \| }

\newcommand{\VEC}[1]{\boldsymbol{#1}}
\newcommand{\MAT}[1]{\boldsymbol{#1}}

\newcommand{\plus}{\raisebox{0\height}{\scalebox{.55}{+}}}

\DeclareMathSymbol{\minus}{\mathbin}{AMSa}{"39}

\usepackage[top=0.7in, bottom=0.7in, left=0.75in, right=0.75in]{geometry} 

\title{Spherical Harmonic Sliced Wasserstein Displacement Interpolation for Acoustic Source and Reflection Density Modeling}
\name{Yuancheng Luo}
\address{NuSpace Audio\\Cambridge, MA, USA}
\begin{document}
\ninept
\maketitle
\begin{abstract}
Spatial room impulse responses (SRIRs) capture directional distributions of acoustic sound-sources and their reflections. However, collecting SRIRs of moving sound-sources remains a challenge, requiring complex interpolations across measurements that account for multi-path spatial-temporal dynamics. This paper investigates the Wasserstein metric and displacement for evaluating interpolated SRIR echo densities in the spherical harmonic domain. We present novel sum-of-magnitude square expansions for efficiently fitting probability density functions, maximizing likelihood, inverse sampling, and computing spherical sliced Wasserstein interpolations. Experiments compare the Wasserstein displacements and metric to linear and geometric interpolations of SRIR image-source densities on a line-path, and demonstrate model-order reduction.
\end{abstract}
\begin{keywords}
Spatial audio, semidefinite programming, polynomials, acoustic propagation, sampling methods
\end{keywords}
\section{Introduction}
\label{SEC:INTRO}

Spatial band-limited room impulse responses (RIRs) have received growing interest in room acoustic modeling and accurate sound-field reproduction. Formats such as higher-order Ambisonics \cite{zotter2019ambisonics, daniel2003further} capture directional characteristics of sound-fields via spherical microphone arrays \cite{jarrett2017theory} and decomposition along the spherical harmonic (SH) basis functions \cite{rafaely2015fundamentals, muller2006spherical}. However, efficient data collection over wide coverage of source and receiver positions remains a challenge due to limits on the number of expensive microphone arrays, and fine-grain discretization of measurement trajectories. Recent efforts therefore move data collection into simulations where RIRs between a stationary source and multiple receivers are simultaneously resolved. For example,  contributions of specular reflections in the classic image-source model (ISM) \cite{allen1979image} are decomposed along SHs and convolved with local receiver directivity \cite{samarasinghe2018spherical, wang2023time, luo2021FSRR}. SRIRs at coordinates between measured or simulated responses require interpolation and analysis of their echo distributions for accurate modeling.

This paper investigates the evolution of SRIR's echo density over the spherical coordinates for a moving source between two points on a line-path. Echo density profiles of real RIRs \cite{abel2006simple, huang2007aspects} can be generalized over spherical coordinates for SRIRs  \cite{tervo2013spatial} and directly evaluated in SH-ISM simulations by tracing image-sources' incident angles and delays. Furthermore, we restrict our study to only the evolution of echo spatial distributions by aggregating the image-source contributions over a SRIR's duration; previous works have used both time and direction-of-arrival of specular reflections \cite{zhao2022interpolating, mckenzie2023source} for interpolating peak-matched SRIRs, and employ non-parametric kernel ridge regression \cite{ribeiro2022region}, and variational Gaussian processes \cite{fernandez2021reconstruction} for spatial-temporal acoustic transfer function interpolation. We motivate the restriction as to efficiently estimate geodesic metrics or shortest-path assignments between acoustic sources and reflections across SRIRs; aggregating a large number of image-sources onto spatial band-limited SH functions yields a tractable measure of echo density over the spherical coordinates. We therefore show that optimal transport \cite{gabriel2019computational} between aggregate echo densities are valid metrics for SRIR interpolation, and are complementary to partial optimal transport of SRIR early-reflection ISM point-clouds \cite{geldert2023interpolation}. Our paper is organized as follows:

Section \ref{SEC:SH_DENSE} introduces our echo density model via sum-of-magnitude square (SOMS) SH expansions and variation of the sum-of-squares (SOS) \cite{lasserre2007sum, roh2006discrete} decomposition of non-negative polynomials. Section \ref{SEC:SH_DENSE_EST} presents novel maximum likelihood density estimation methods of echo distributions, and section \ref{SEC:SH_DENSE:SAMPLING} derives efficient inverse transform sampling schemes.
Section \ref{SEC:SH_DENSE:OPT_TRANS} extends the sampling methods to the spherical sliced Wasserstein (SSW) \cite{quellmalz2023sliced, bonet2023spherical} displacement interpolations in the SH domain via our non-negative least-squares (NNLS) \cite{lawson1974linear} and semidefinite programming (SDP) \cite{vandenberghe1996semidefinite,cvx,parrilo2003semidefinite} optimization. 
Section \ref{SEC:EXP} reports experimental results of interpolating acoustic reflection densities of SH-ISM methods across a moving acoustic source path. 
Code is available online\footnote{\url{https://github.com/yluo1/SH-IRT}/}.

\section{Spherical Harmonic Density Function}
\label{SEC:SH_DENSE}

\begin{figure*}[tb]
\centering 
    \subfloat[Kernel log-likelihood: $-222.5$ \label{FIG:DENSITY:PROJ}]{%
        \includegraphics[width=0.32\textwidth]{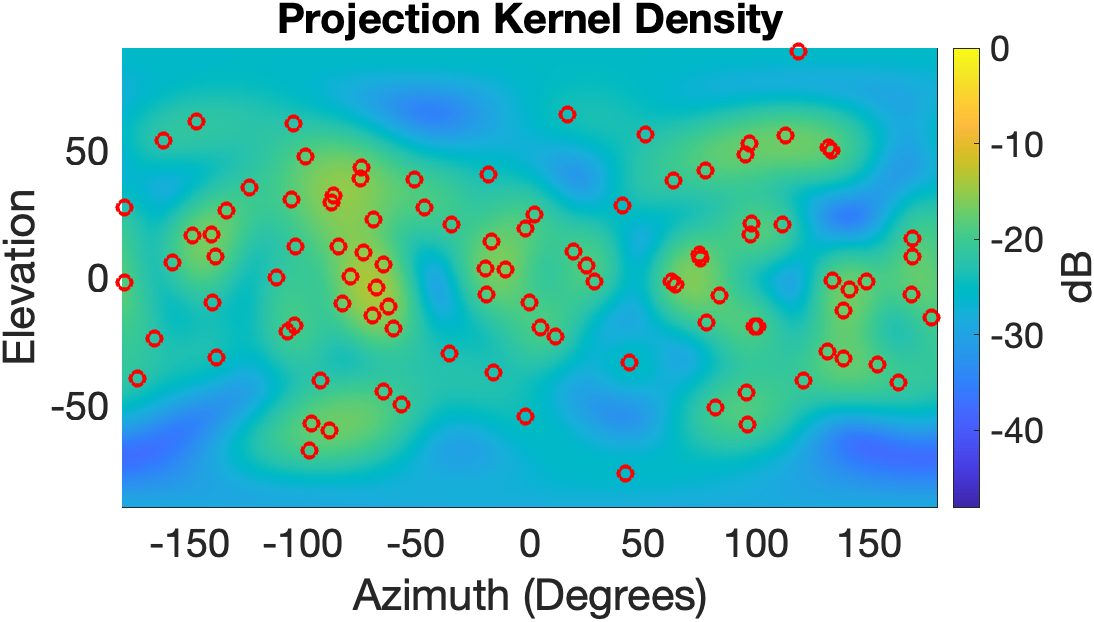}%
    }
    \hspace{1pt}
    \subfloat[Maximum spectral log-likelihood: $-209.7$\label{FIG:DENSITY_MAX_SPECTRA}]{%
        \includegraphics[width=0.323\textwidth]{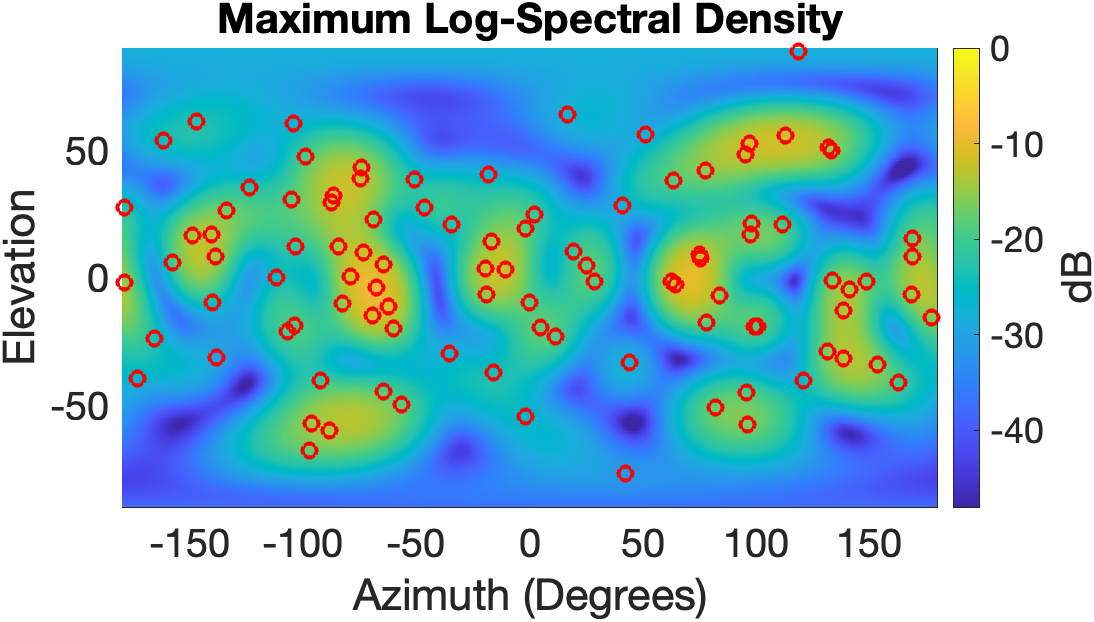}%
    }
        \hspace{1pt}
   \subfloat[Maximum log-likelihood: $-192.9$\label{FIG:DENSITY_MAX_LIKE}]{%
        \includegraphics[width=0.32\textwidth]{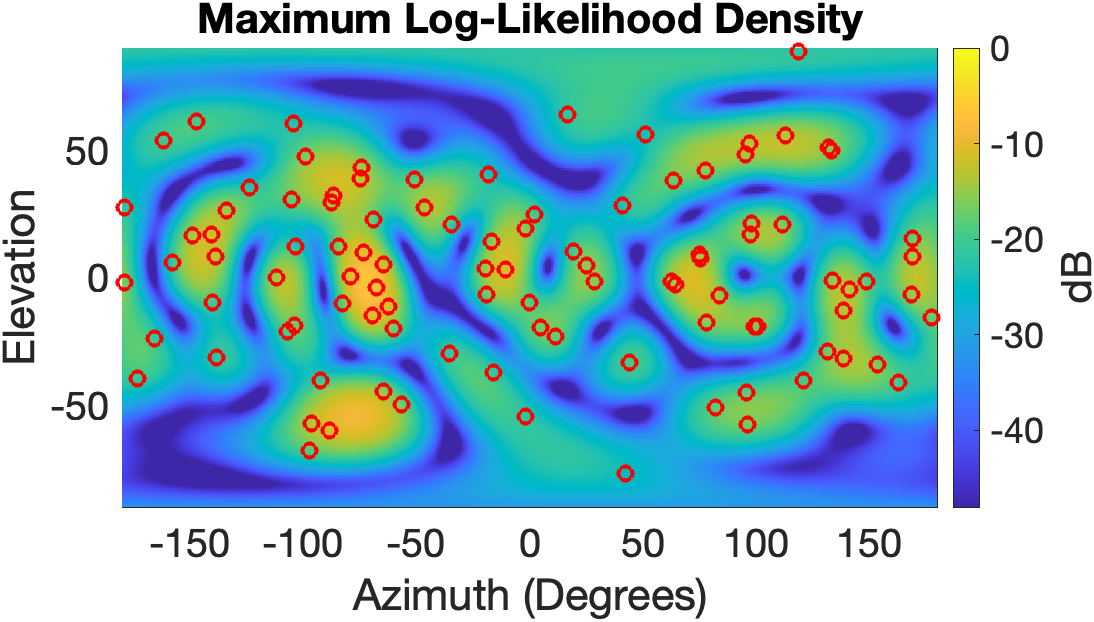}%
    }
      \caption{\label{FIG:DENSITY}SH density functions (max-order $L_C = 6$) are estimated over directions ($N=100$ red circles) sampled from a uniform distribution over spherical coordinates. The projection kernels, spectral, and the maximum likelihood optimized densities span the range of bias-variance and computational cost trade-offs.  For comparison, a uniform density with SH coefficients $\VEC{C} = \VEC{e}_1$ (standard basis) obtains a log-likelihood of $-253.1$.}
\end{figure*}

Let the function $f_C(\theta, \phi)$ over the spherical coordinates $(\theta, \phi)$ be expanded along finite max-order $L_C$ SH bases $Y(\theta, \phi)$ given by
\begin{equation} \label{EQ:SH_DENSITY_RE}
\displaystyle
\begin{split}
f_{\VEC{C}}(\theta, \phi) & = \sum_{l = 0}^{L_C} \sum_{m= \minus l}^l  Y_l^m(\theta, \phi) C_l^m, \qquad \textrm{SH Expansion}  \\
Y_l^m(\theta, \phi) & = \sum_{l=0}^{L_C} \sum_{m = \minus l}^{l} \sqrt{\frac{(2l + 1)}{4 \pi } \frac{(l-m)!}{(l+m)!}} \,  P_l^m \PAREN{\cos \theta } e^{i m \phi }, 
\end{split}
\end{equation}
where $\VEC{C} = \BRAK{C_0^0, C_{\minus 1}^1, \hdots C_{L_C}^{L_C} }^T \in \field{C}^{N_C \times 1}$ is the vector of SH coefficients for $N_C = (L_D + 1)^2$, and $P_l^m(\cos \theta)$ the associated Legendre polynomial functions of degree $l$ and order $m$.
We can construct a probability density function that is non-negative with SH expansions by constraining it to be SOMS functions expressed by $f_{\VEC{D}}(\theta, \phi) = \sum_{s=1}^{S} \ABS{f_{\VEC{C}_s}(\theta, \phi)}^2$ of $S$ number of expansions following
\begin{equation} \label{EQ:SH_DENSITY_NON_NEG}
\displaystyle
\begin{split}
f_{\VEC{D}}(\theta, \phi) & = \sum_{l = 0}^{L_D} \sum_{m= \minus l}^l  Y_l^m(\theta, \phi) D_l^m   =
\sum_{s=1}^S f_{\VEC{C}_s}(\theta, \phi) f_{\VEC{C}_s}^*(\theta, \phi), \\
\VEC{D} & = \sum_{s=1}^S \VEC{C}_s \diamond \tilde{\VEC{C}}_{s}  = \BRAK{D_0^0, D_{\minus 1}^1, \hdots D_{L_D}^{L_D} }^T \in \field{C}^{N_D \times 1},
\end{split}
\end{equation}
which sums over products of SH expansion $f_{C_s}(\theta, \phi)$ and its conjugate $f_{C_s}^*(\theta, \phi)$. The expansion coefficients $\VEC{D}$ are computed via SH product $\diamond$ of SH conjugates $\tilde{\VEC{C}}$ operators as defined in \cite{luo2021spherical} in twice the expansion max-order $L_D = 2 L_C$, for $N_D = (L_D + 1)^2$.
We impose the unity summation constraint on density $f_D(\theta, \phi)$ via normalization:
\vspace{-1ex}
\begin{equation} \label{EQ:SH_DENSITY_UNITY}
\displaystyle
\begin{split}
\int_{\theta = 0}^{\pi} \int_{\phi = 0}^{2 \pi} \sum_{l = 0}^{L_D} \sum_{m= \minus l}^l  Y_l^m(\theta, \phi) D_l^m  \sin\theta \, d\theta \, d\phi & = 1  \\
\Leftrightarrow \quad D_0^0 = \frac{1}{2 \sqrt{\pi}} \quad \Leftrightarrow \quad
\sum_{s = 1}^S \VEC{C}^H_s \VEC{C}_s & = 1,
\end{split}
\end{equation}
where the orthogonality of SH bases requires that the $0^{th}$ degree coefficient of $\VEC{D}$ is normalized to a constant, and is also satisfied if the total sum-of-products $\VEC{C}_s$ and Hermitian transpose $\VEC{C}_s^H$ is unity.

\subsection{Density Function Estimation}
\label{SEC:SH_DENSE_EST}

Let the log-likelihood function $\ell (\VEC{D} \, | \, \VEC{\theta}, \VEC{\phi} )$ of the set of observed reflection directions $\CBRAK{\VEC{\theta}, \VEC{\phi} \in \field{R}^N}$ for density $f_{\VEC{D}}(\theta, \phi)$ be given by
\begin{equation} \label{EQ:SH_DENSITY_LIKELIHOOD}
\displaystyle
\begin{split}
\ell (\VEC{D})  & = \sum_{n=1}^N \log \PAREN{ \sum_{s=1}^S {  \PAREN{ \VEC{C}_s^T  \VEC{Y}_{\theta_n \phi_n} }^* \PAREN{ \VEC{Y}_{\theta_n \phi_n}^T \VEC{C}_s }}} \\
& = \sum_{n=1}^N \log \PAREN{ \trace{ \MAT{Q} \PAREN{  \VEC{Y}_{\theta_n \phi_n}   \VEC{Y}_{\theta_n \phi_n}^H }   } },
\quad \MAT{Q} = \sum_{s = 1}^S \VEC{C}_s \VEC{C}_s^H, 
\end{split}
\raisetag{11ex}
\end{equation}
where $\VEC{Y}_{\theta_n \phi_n} = \BRAK{Y_0^0(\theta_n, \phi_n), Y_{\minus 1}^1(\theta_n, \phi_n), \hdots, Y_{L_C}^{L_C}(\theta_n, \phi_n) }^T \in \field{C}^{N_C \times 1}$ is a vector containing the SH function evaluations at spherical coordinate $(\theta_n, \phi_n)$, and $\MAT{Y}_{\VEC{\theta}\VEC{\phi}} = \BRAK{\VEC{Y}_{\theta_1 \phi_1}, \hdots,  \VEC{Y}_{\theta_N \phi_N} } \in \field{C}^{N_C \times N}$ is the column matrix of evaluations across all observed directions.
Applying the matrix trace operator $\trace{*}$, we can express the log-likelihood in terms of the empirical covariance matrix $\MAT{Q} \in \field{C}^{N_C \times N_C}$ between SH bases of the unknown SH expansion coefficient vectors $\VEC{C}_s$. Thus, we optimize over positive semidefinite constrained matrices $\MAT{Q}$.

Consider maximizing log-likelihood Eq. \eqref{EQ:SH_DENSITY_LIKELIHOOD} by transforming the complex SH bases and coefficients into their real forms given by $\bar{\VEC{Y}}_{\theta_n \phi_n} = \MAT{U} \VEC{Y}_{\theta_n \phi_n}$, and $\bar{\VEC{C}}_s = \MAT{U} \VEC{C}_s$ respectively, via the complex to real SH transformation matrix $\MAT{U} \in \field{C}^{N_C \times N_C}$ as defined in \cite{politis2024gaunt}.
Its sum-of-log objective function is concave which ensures that the local maximum is the global maximum when the feasible space is convex.
We can therefore maximize the equivalent log-likelihood function of real variables via the following SOS-SDP in CVX \cite{cvx}:
\begin{equation} \label{EQ:SH_DENSITY_LIKELIHOOD_OPT}
\displaystyle
\begin{split}
\bar{\VEC{Q}}_* & = \argmax{\bar{\MAT{Q}}} \ell \PAREN{\bar{\VEC{Q}} } \qquad  \textrm{s.t.} \qquad 
 \bar{\MAT{Q}} \succeq 0, \quad \trace{\bar{\VEC{Q}}} = 1, \\ 
 \ell \PAREN{\bar{\VEC{Q}} } & = \sum_{n=1}^N \log  \PAREN{  \trace{ \bar{\MAT{Q}} \PAREN{  \bar{\VEC{Y}}_{\theta_n \phi_n} \bar{\VEC{Y}}_{\theta_n \phi_n}^T } } },  \quad 
 \bar{\MAT{Q}}_*  = \MAT{V} \MAT{\Sigma} \MAT{V}^T,
\end{split}
\raisetag{12.5ex}
\end{equation}
where $\bar{\MAT{Q}} \in \field{R}^{N_C \times N_C}$ is constrained to be positive semidefinite.
The column-matrix $\bar{\MAT{C}}_* = \MAT{V} \MAT{\Sigma}^{\frac{1}{2}}  \in \field{R}^{N_C \times N_C}$ contains the SH expansion coefficients recovered from the eigendecomposition of the SDP solution $\bar{\MAT{Q}}_*$; non-zero columns $\bar{\VEC{C}}_{*s}$ correspond to eigenvectors weighted by the square-root of non-zero eigenvalues in $\MAT{V}, \, \MAT{\Sigma} \in \field{R}^{N_C \times N_C}$ respectively. 
Furthermore, $\bar{\VEC{C}}_{*s}$ satisfies the normalization condition in Eq. \eqref{EQ:SH_DENSITY_UNITY} following the unity trace equality constraint of Eq. \eqref{EQ:SH_DENSITY_LIKELIHOOD_OPT} as  $\trace{\bar{\MAT{Q}}_*} = \trace{\MAT{\Sigma}} = 1$.
The final SOS SH expansion $\bar{\MAT{D}}_* = \sum_{n=1}^{N_C} \bar{\VEC{C}}_{*n} \diamond \bar{\VEC{C}}_{*n} $ follows the sum of SH product $\diamond$ operators over typically sparse set of columns of $\bar{\MAT{C}}$. We now compare the SDP solution to alternative density estimation methods:

Let us specify a baseline kernel density estimator via the normalized SH projection  $\VEC{C}_{n}  =  \VEC{Y}_{\theta_n \phi_n}^* /  \sum_{n=1}^N \VEC{Y}_{\theta_n \phi_n}^{H} \VEC{Y}_{\theta_n \phi_n} $ whereby projections at each observed spherical coordinate $\theta_n, \phi_n$ yield a finite-order SH expansion of a Dirac function over the sphere via the Legendre addition theorem. The resulting expansion is an isotropic kernel function over the spherical coordinates given by
\begin{equation} \label{EQ:SH_PROJ}
\displaystyle
\begin{split}
f_{\VEC{C}_n} (\theta, \phi)  & \propto  \sum_{l = 0}^{L_C} \sum_{m= \minus l}^l  Y_l^m(\theta, \phi)  Y_l^{m*}(\theta_n, \phi_n),  \\
\end{split}
\end{equation}
which has a kernel bandwidth proportional to the maximum SH order, and is equivalent to the Ambisonics encoding of a unit pulse panned to $\theta_n, \phi_n$.
It is well-known that spatial aliasing of finite-order expansions introduces spurious ripples over the spherical coordinates, which results in over-smoothed SOMS densities as shown in Fig. \ref{FIG:DENSITY:PROJ}. We can increase the estimator's data likelihood by substituting $\VEC{C}_n$ into $\VEC{C}_s$ within the log-summation Eq. \eqref{EQ:SH_DENSITY_LIKELIHOOD}, and decompose the empirical covariance matrix  $\MAT{Q} = \sum_{n = 1}^N \VEC{C}_n \VEC{C}_n^H = \MAT{V} \MAT{\Sigma} \MAT{V}^H$ along its eigenvectors. The trace terms in the log-likelihood can be expressed as inner-products $\trace{ \PAREN{ \MAT{V}^T  \VEC{Y}_{\theta_n \phi_n}   \VEC{Y}_{\theta_n \phi_n}^H \MAT{V} } \MAT{\Sigma} } = \VEC{\mu}_n^T \VEC{\lambda}$ where $\VEC{\mu}_n, \, \VEC{\lambda} \in \field{R}^{N_C \times 1}$ are diagonal entrants of the eigenvector projections $\PAREN{\MAT{V}^T  \VEC{Y}_{\theta_n \phi_n}   \VEC{Y}_{\theta_n \phi_n}^H \MAT{V}}$ and the eigenvalue matrix $\MAT{\Sigma}$ respectively. Therefore, we can maximize the log-likelihood objective w.r.t. spectral weights $\VEC{\lambda}$ of the eigenvector projections in the following convex program with unity summation constraints:
\begin{equation} \label{EQ:SH_DENSITY_LIKELIHOOD_OPT_SPECTRA}
\displaystyle
\begin{split}
\VEC{\lambda}_* & = \argmax{\VEC{\lambda}}   \sum_{n=1}^N \log  \PAREN{   \VEC{\mu}_n^T \VEC{\lambda} }  \quad  \textrm{s.t.} \quad 
\VEC{\lambda} \geq \VEC{0}, \quad  \VEC{\lambda}^T \VEC{1} =  1, 
\end{split}
\end{equation}
which reduces the number of variables to $N_C$  compared to $(N_C / 2)^2$ in the SDP formulation of Eq. \eqref{EQ:SH_DENSITY_LIKELIHOOD_OPT}.
The resulting solutions sharpen the density estimate as shown in Fig. \ref{FIG:DENSITY_MAX_SPECTRA}, increasing the data likelihood over the kernel density. Note that maximum log-likelihood solutions risk over-fitting to the observed samples shown in Fig. \ref{FIG:DENSITY_MAX_LIKE}, as the underlying sample distribution is uniform. Thus, the log-spectral densities of Eq. \eqref{EQ:SH_DENSITY_LIKELIHOOD_OPT_SPECTRA} balance both bias-variance and computational costs compared to kernel density and maximum likelihood estimators. 
We now show how to efficiently sample directions from our SOMS densities via inverse transforms \cite{devroye1986sample}.

\subsection{Inverse Transform Sampling}
\label{SEC:SH_DENSE:SAMPLING}

\begin{figure}[tb]
\centering
  \includegraphics[width=0.385\columnwidth]{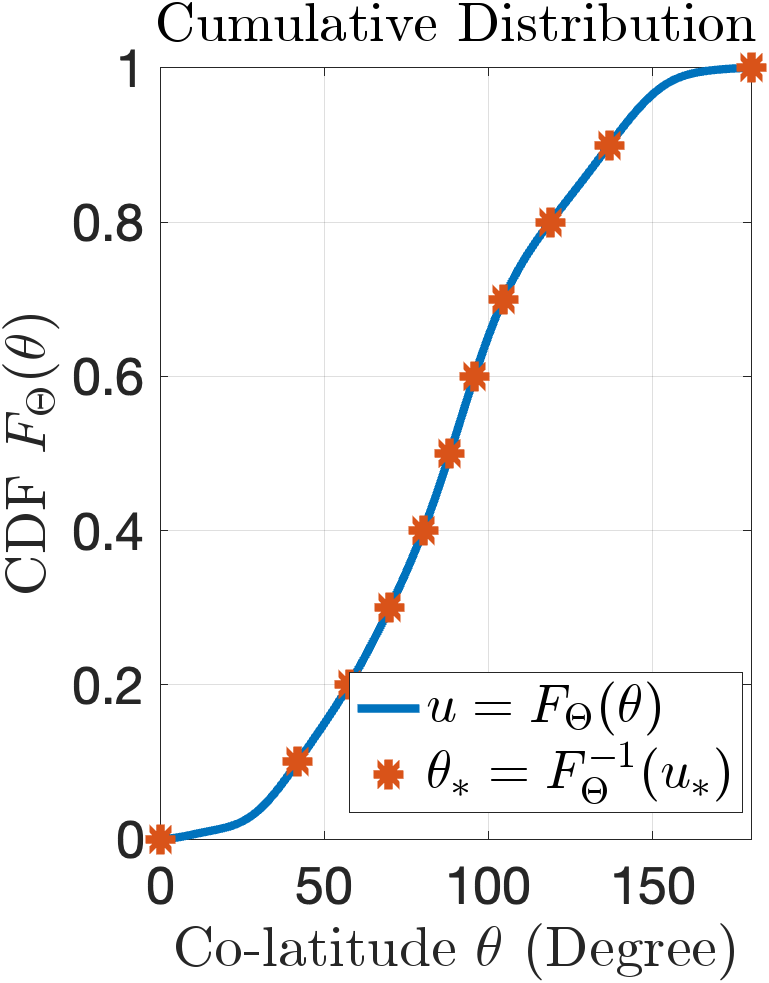}
   \includegraphics[width=0.485\columnwidth]{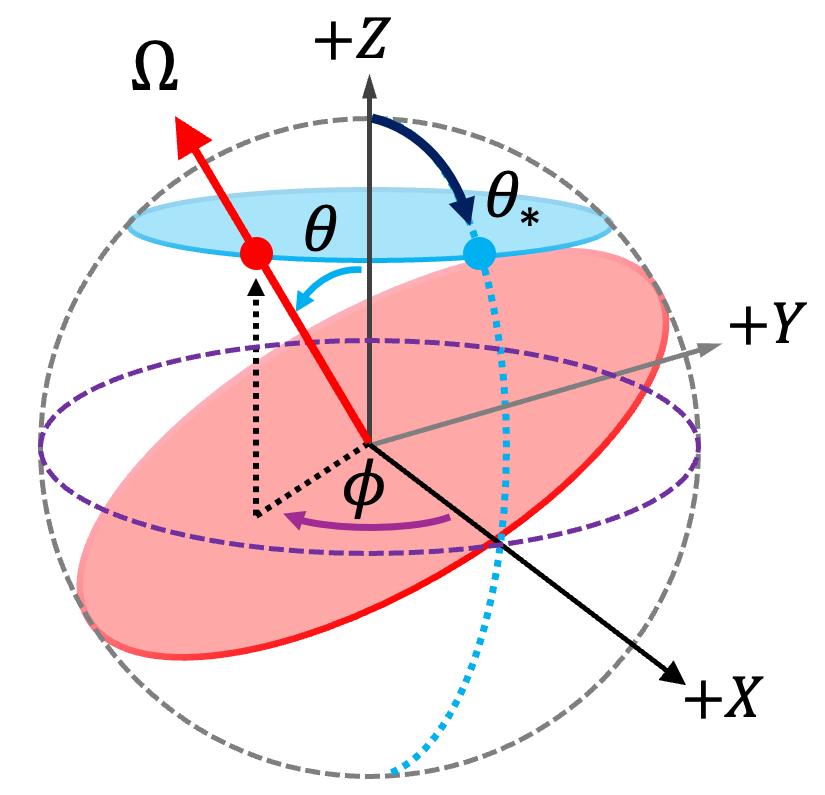}
      \caption{\label{FIG:SH_COORD_CDF}The sample marginal CDF $F_{\Theta}(\theta_*)$ for the maximum log-likelihood SH expansion in Fig. \ref{FIG:DENSITY_MAX_LIKE} integrates the PDF over the cap $(0 \leq \Theta \leq \theta_*, \, 0 \leq \Phi < 2 \pi )$ along the meridian semi-circle. }
\end{figure}

Let the joint cumulative distribution function (CDF) of the density function's random variables (co-latitude $\Theta$, azimuth $\Phi$) be given by 
\begin{equation} \label{EQ:JOINT_CDF}
\displaystyle
\begin{split}
F_{\Theta \Phi}(\theta, \phi) & = \sum_{l=0}^{L_D} \sum_{m = \minus l}^{l} \sqrt{\frac{(2l + 1)}{4 \pi } \frac{(l-m)!}{(l+m)!}} \,  D_l^m \\
  & \times  \int_{\Theta=0}^{\theta} \int_{\Phi = 0}^{\phi}  P_l^m \PAREN{\cos\Theta } e^{i m \Phi } \sin \Theta \, d\Theta \, d\Phi , 
\end{split}
\end{equation}
where the marginal CDF $F_{\Theta}(\theta) = F_{\Theta \Phi}(\theta, 2 \pi) $ is analytic following
\begin{equation} \label{EQ:MARGIN_CDF}
\displaystyle
\begin{split}
F_{\Theta}(\theta) & = \sum_{l=0}^{L_D} \sqrt{\frac{\pi}{ 2l + 1 }  }  \,  D_l^0  
   \left \{ \begin{array}{cc} 
    1 -  \cos\theta , & l = 0  \\
    P_{l \minus 1} \PAREN{\cos\theta } -  P_{l \plus 1} \PAREN{\cos\theta } , & l > 0  \\
    \end{array} \right . . 
\end{split}
\end{equation}
For co-latitude, we sample $\Theta$ via the inverse transform of the marginal CDF by first sampling $u_* \sim \mathcal{U}(0, 1)$ from a uniform distribution and then efficiently computing the CDF's inverse $\theta_* = F_{\Theta}^{\minus 1}(u_*)$. We consider the bracketed Newton-Raphson iterations $\theta_{k+1} = \theta_{k} - (F_{\Theta}(\theta_k) - u_*) /  F'_{\Theta}(\theta_k)$, which quickly converge as bounds around the shifted monotone CDF tighten around the single root in the interval $\BRAK{0, \pi}$. The CDFs have smooth derivatives over the Legendre polynomials following
\begin{equation} \label{EQ:MARGIN_CDF_DERIV}
\displaystyle
\begin{split}
\frac{d F_{\Theta}(\theta)}{d \theta } & = \sum_{l=0}^{L_D}  \sqrt{ (2l + 1) \pi  } \,  D_l^0  P_{l} \PAREN{\cos\theta } \,  \sin \theta,
\end{split}
\end{equation}
where iterations over the domain of $\cos \theta \in \BRAK{-1, 1}$ avoids trigonometric function evaluations.
The CDFs are therefore smooth as seen in Fig. \ref{FIG:SH_COORD_CDF} but have small derivatives near the poles. Thus, we switch to the bisection method when Newton-Rapshon iterations falls outside the bracket.

For azimuth, we sample $\Phi$ from the inverse transform of the conditional CDF $F_{\Phi}(\phi)$ given $\theta_*$. The former is also analytic and whose derivative is proportional to the density's SH expansion following
\begin{equation} \label{EQ:COND_CDF}
\displaystyle
\begin{split}
F_{\Phi}(\phi) &  = \frac{ \int_{\Phi = 0}^{\phi} f_{\VEC{D}}(\theta_*, \Phi)  d \Phi }{ F_{\VEC{D}}(\theta_*) }, \quad
\frac{d F_{\Phi}(\phi)}{d \phi}  = \frac{ f_{\VEC{D}}(\theta_*, \phi) }{ F_{\VEC{D} }(\theta_*) }, 
\end{split}
\end{equation}
where the density term in the numerator and the normalization term $F_{\VEC{D}}(\theta_*) = \int_{\Phi = 0}^{2 \pi } f_{\VEC{D}}(\theta_*, \Phi)  d \Phi $ in the denominators are expressed by
\begin{equation} \label{EQ:COND_CDF_TERMS}
\displaystyle
\begin{split}
 \int_{\Phi = 0}^{\phi} f_{\VEC{D}}(\theta_*, \Phi)  d \Phi 
& = \sum_{l=0}^{L_D} \sum_{m =  \minus l}^{l} \sqrt{\frac{(2l + 1)}{4 \pi } \frac{(l-m)!}{(l+m)!}} \,  D_l^m  \\
& \times  P_l^m \PAREN{\cos\theta_* } 
 \left \{ \begin{array}{cc} 
 \phi, & m = 0 \\[2.5pt]
 \frac{i \PAREN{1 - e^{i m \phi}} }{m}, & m \neq 0
   \end{array} \right . ,  \\
    \int_{\Phi = 0}^{2 \pi} f_{\VEC{D}}(\theta_*, \Phi)  d \Phi
   & =    \sum_{l=0}^{L_D}  \sqrt{(2l + 1) \pi}  \,   D_l^0 P_l \PAREN{\cos\theta_* }.
\end{split}
\raisetag{5ex}
\end{equation}
Therefore, we can sample the azimuth coordinate $\phi_*$ paired with $\theta_*$, and a second sample $u_*$ drawn from the uniform distribution, by computing the inverse conditional CDF $\phi_* = F_{\Phi | \theta_*}^{\minus 1} (u_*)$ via analogous bracketed Newton-Rapshon iterations: $\phi_{k+1} = \phi_{k} - (F_{\Phi | \theta_*}(\phi_k) - u_*) /  F'_{\Phi | \theta_*}(\phi_k)$ in the interval of $\phi \in \BRAK{0, 2 \pi}$, where
pre-computing $P_l^m (\cos \theta_*)$ further reduces costs per iteration. 
We now apply our $1$-dimensional inverse transforms to projections of the spherical Wasserstein metric.

\section{Spherical Sliced Wasserstein Interpolation}
\label{SEC:SH_DENSE:OPT_TRANS}

\begin{figure*}[tb]
\centering
 \captionsetup[subfloat]{farskip=2pt, captionskip=1pt}
    \subfloat{%
        \includegraphics[width=0.2025\textwidth]{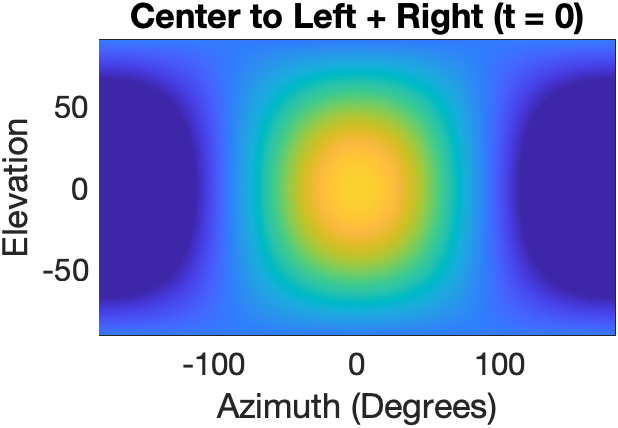}%
    }
    \hspace{1pt}
    \subfloat{%
        \includegraphics[width=0.18\textwidth]{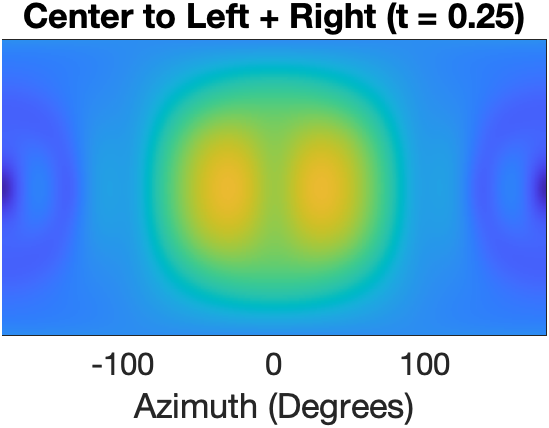}%
    }
        \hspace{1pt}
   \subfloat{%
        \includegraphics[width=0.18\textwidth]{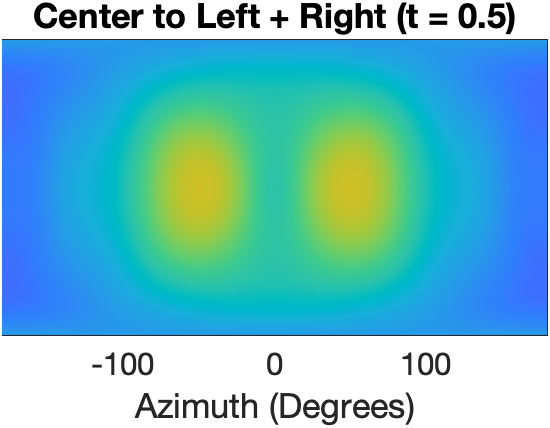}%
    }
            \hspace{1pt}
   \subfloat{%
        \includegraphics[width=0.18\textwidth]{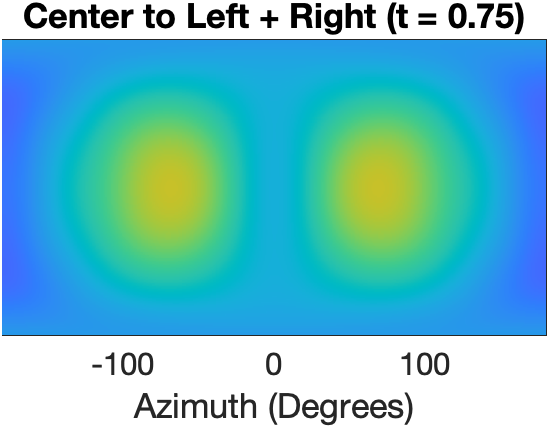}%
    }
      \hspace{1pt}
   \subfloat{%
        \includegraphics[width=0.18\textwidth]{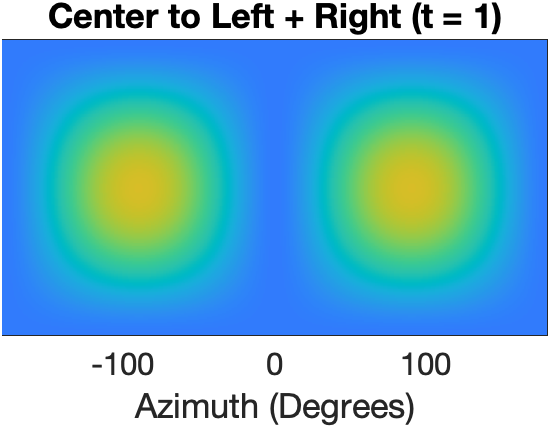}%
    } \\
    \subfloat{%
        \includegraphics[width=0.2025\textwidth]{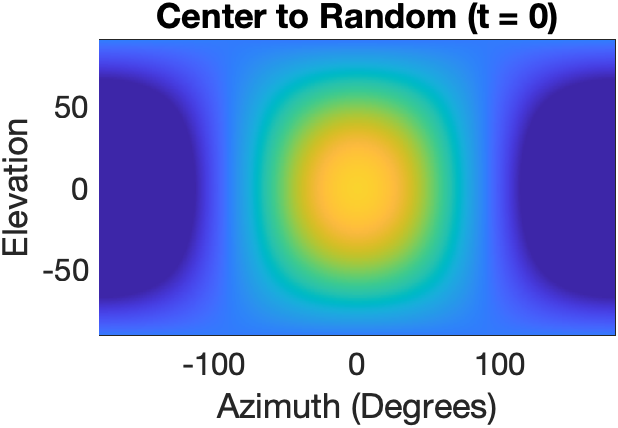}%
    }
    \hspace{1pt}
    \subfloat{%
        \includegraphics[width=0.18\textwidth]{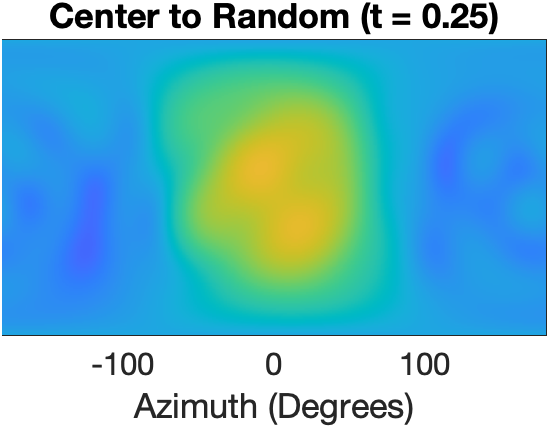}%
    }
        \hspace{1pt}
   \subfloat{%
        \includegraphics[width=0.18\textwidth]{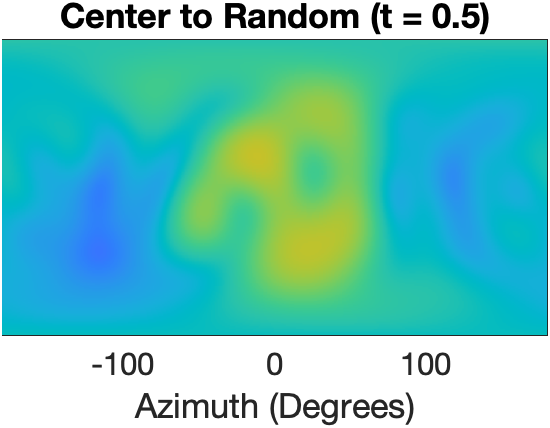}%
    }
            \hspace{1pt}
   \subfloat{%
        \includegraphics[width=0.18\textwidth]{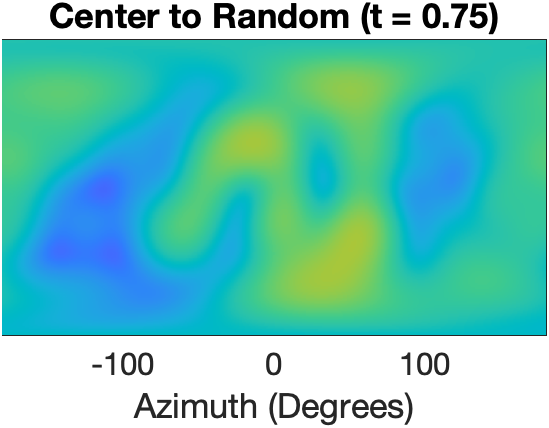}%
    }
      \hspace{1pt}
   \subfloat{%
        \includegraphics[width=0.18\textwidth]{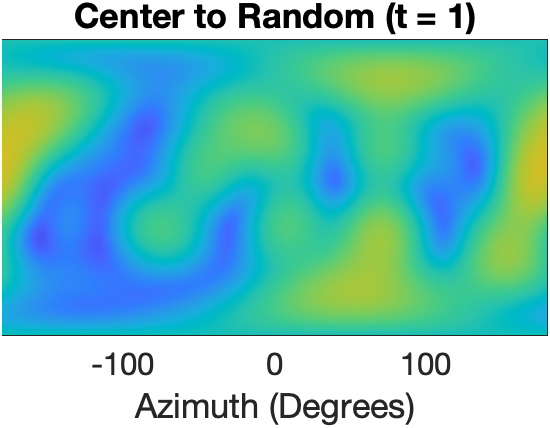}%
    }
      \caption{\label{FIG:DISP_INTERP}Wasserstein interpolations via Eqs. \eqref{EQ:WASS_SLICE_NNLS}, \eqref{EQ:WASS_SLICE_SDP} smoothly transport the SH expansion of squared exponential kernel (max-order $L=8$, $\ell = 0.75$) of chordal distances \cite{luo2021spherical} density (left column) to bifurcated and randomized densities (right column) across time-steps $t \in \CBRAK{0, \, 0.25, \, 0.5, \, 0.75, \, 1}$.}
\end{figure*}

\begin{figure*}[htb]
\centering
 \captionsetup[subfloat]{farskip=6pt, captionskip=1pt}
    \hspace{10pt}
 \subfloat{
         \includegraphics[width=0.2325\textwidth]{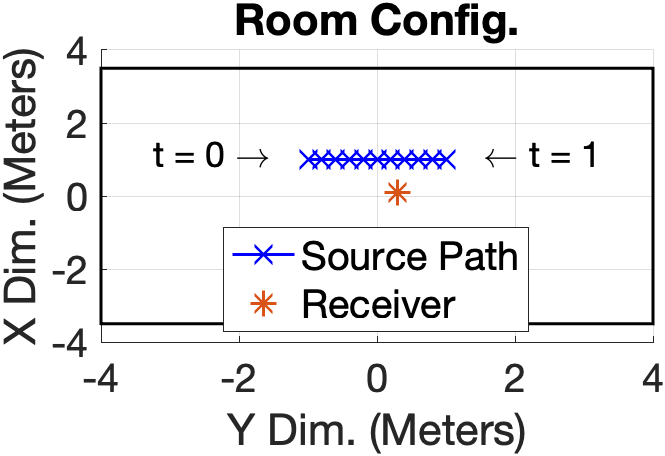}%
 } 
   \hspace{10pt}
   \subfloat{%
        \includegraphics[width=0.255\textwidth]{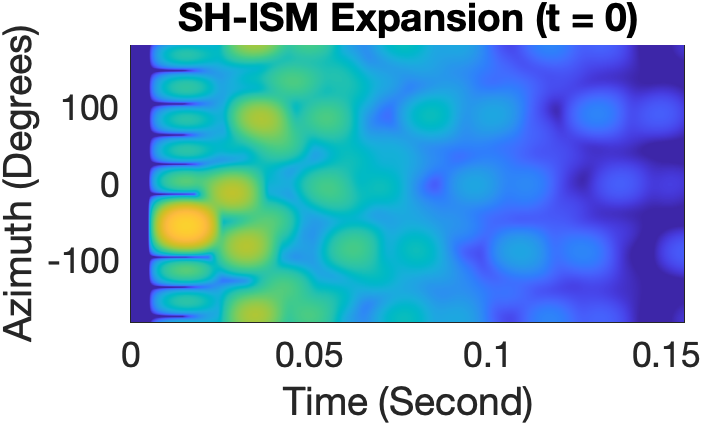}%
    }
   \subfloat{%
        \includegraphics[width=0.23\textwidth]{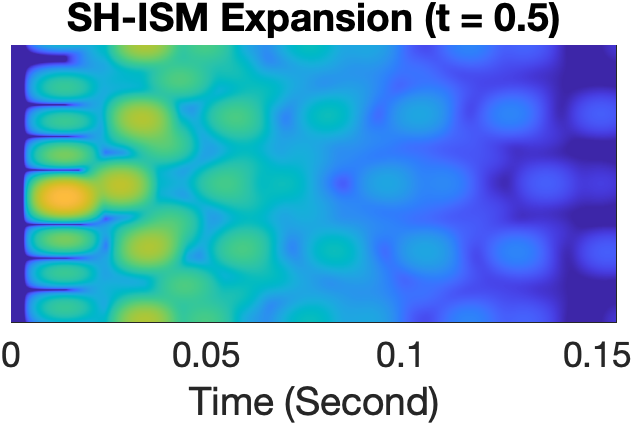}%
    }
   \subfloat{%
        \includegraphics[width=0.23\textwidth]{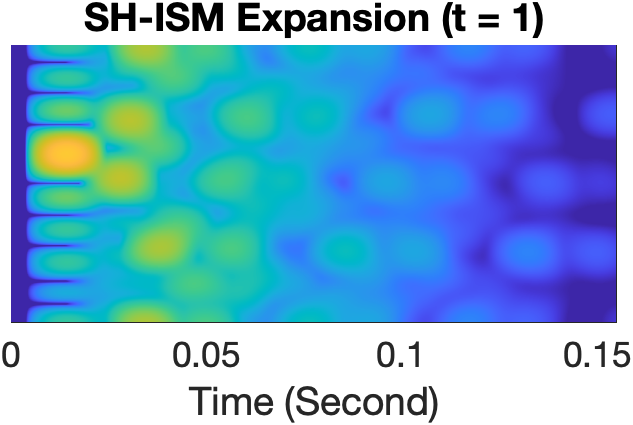}%
    } \\
   \subfloat{%
        \includegraphics[width=0.29\textwidth]{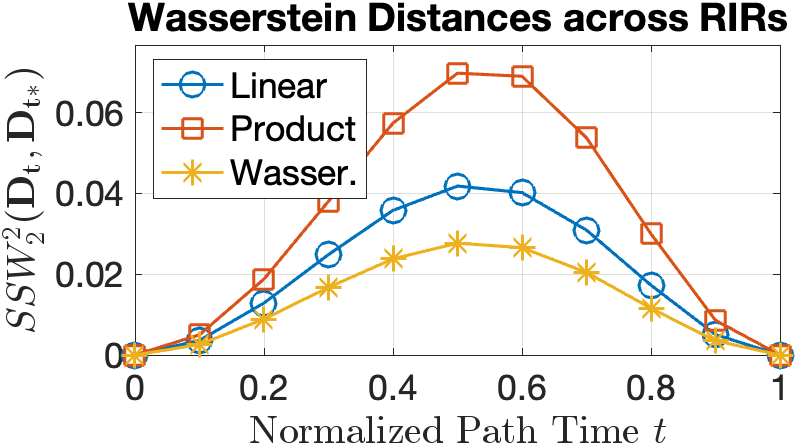}%
    }
   \subfloat{%
        \includegraphics[width=0.2475\textwidth]{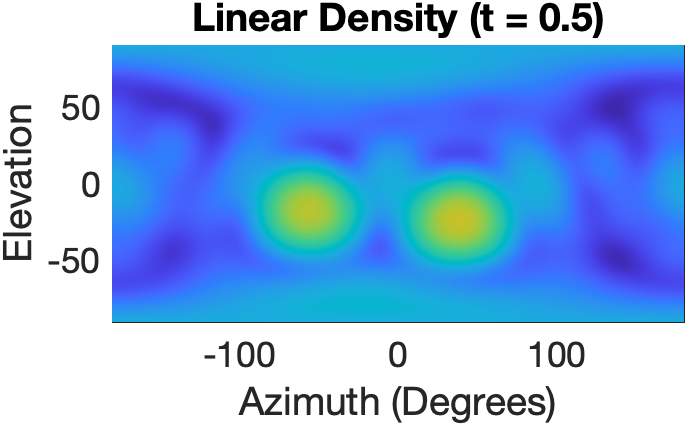}%
    }
            \hspace{1pt}
   \subfloat{%
        \includegraphics[width=0.22\textwidth]{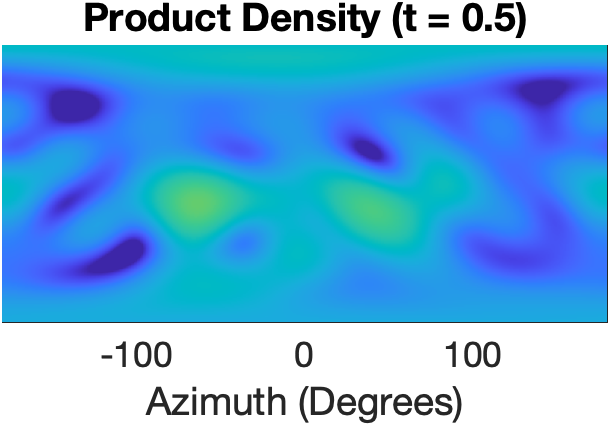}%
    }
      \hspace{1pt}
   \subfloat{%
        \includegraphics[width=0.22\textwidth]{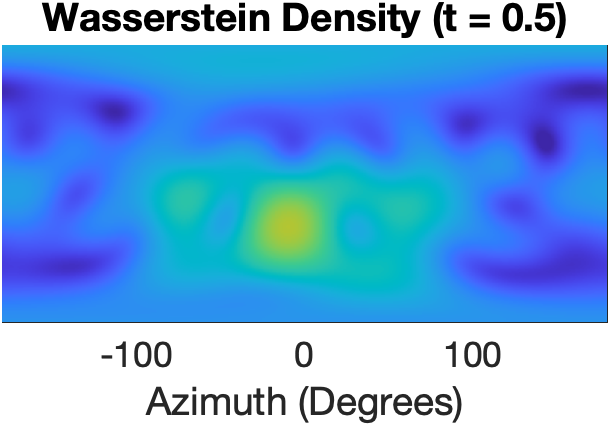}%
    }
      \caption{\label{FIG:RIR_EXP}Ground-truth RIRs of an omni-directional sound-source to receiver are computed over a left-to-right line-path indexed by time-stamp $t$, and their direct and image-sources expanded over the SH domain (top row). Source densities $\VEC{D}_{t*}$ aggregate direct and image-source SH expansions over the RIR's duration at each $t$. Linear, product, and Wasserstein interpolated densities $\VEC{D}_{t}$ at end-point supports ($t=0$, $t=1$) are compared to ground-truth densities $\VEC{D}_{t*}$, and shown for $t=0.5$ (bottom row). Wasserstein interpolations achieve the lowest SSW distances from the ground-truth across the entire path.}
\end{figure*}

The Wasserstein metric measures the minimum transport cost across all possible coupling between two input densities. For SH densities, the spherical sliced 2-Wasserstein distance $SSW_{2}^2(\VEC{D}_0, \VEC{D}_1)$ between densities $f_{\VEC{D}_0}(\theta, \phi)$ and $f_{\VEC{D}_1}(\theta, \phi)$ are attractive as they can be efficiently computed from projections onto uniformly distributed 2-dimensional great circles on the sphere. Furthermore, projections onto great semi-circles induce 1-dimensional Wasserstein distances $W_{2}^2(\VEC{D}_0, \VEC{D}_1)$ between marginal CDFs of Eq. \eqref{EQ:MARGIN_CDF} for SH density expansion coefficients $\VEC{D}_0$, $\VEC{D}_1$, which can be computed by integrating the square difference of inverse CDFs over the meridian in Fig. \ref{FIG:SH_COORD_CDF}. The SSW distance is therefore given by
\begin{equation} \label{EQ:WASS2_1D}
\displaystyle
\begin{split}
SSW_2^2(\VEC{D}_0, \VEC{D}_1) & = \int_{\Omega} W_2^2( \MAT{R}_{\Omega} \VEC{D}_0, \, \MAT{R}_{\Omega} \VEC{D}_1) \, d \MAT{R}_{\Omega}, \\
W_{2}^2(\VEC{D}_0, \VEC{D}_1 )  & =  \int_{u = 0}^{1}  \ABS{ F_{\Theta}^{\minus 1}(u \, | \, \VEC{D}_0) - F_{\Theta}^{\minus 1}(u \, | \, \VEC{D}_1) }^2  \,  d u,
\end{split}
\raisetag{10ex}
\end{equation}
where matrices $\MAT{R}_{\Omega} \in \field{C}^{N_D \times N_D}$ are SH rotations whose first-degree $l = 1$ are uniformly sampled from the rotation group $SO(3)$ as in \cite{quellmalz2024parallelly}, and indexed by each great circle's normal direction $\Omega \in \field{R}^3$ in Fig. \ref{FIG:SH_COORD_CDF}. We sample normal directions from the spherical Fibonnaci spiral \cite{sisouk:hal-04932591, hannay2004fibonacci} of $8 N_D$ points on the unit-sphere.
Each 1-dimensional $W_{2}^2( \MAT{R}_{\Omega} \VEC{D}_0, \, \MAT{R}_{\Omega} \VEC{D}_1 )$ integral can be efficiently computed via the inverse transform, or by piece-wise linear approximation of the inverse CDFs over the domain $u_i = F_{\Theta}(\theta_i)$ and range $\theta_i$ at uniform radian supports between $\BRAK{0, \pi}$.

For computing sliced displacement interpolations at normalized time $t \, | \, 0 \leq t \leq 1$ between density $f_{\VEC{D}_0}(\theta, \phi)$ and $f_{\VEC{D}_1}(\theta, \phi)$ located on times $t = 0$ and $t = 1$ respectively, inverse CDFs of the 1-dimensional projections are linearly interpolated to yield the desired inverse CDFs given by
\begin{equation} \label{EQ:WASS_SLICE_INTERP}
\displaystyle
\begin{split}
\theta(u, \Omega, t) = (1 - t) \,  F_{\Theta}^{\minus 1} \PAREN{u \, | \,  \MAT{R}_{\Omega} \VEC{D}_0 } + t \, F_{\Theta}^{\minus 1} \PAREN{u \, | \,    \MAT{R}_{\Omega} \VEC{D}_1}.
\end{split}
\end{equation}
We consider fitting the marginal CDFs $F_{\Theta}(\theta(u, \Omega, t) \, | \, \MAT{R}_{\Omega} \VEC{D} )$ in Eq. \eqref{EQ:MARGIN_CDF} of the unknown SH expansion coefficients $\VEC{D}$ to uniformly discretized quantiles $\VEC{u} = \BRAK{u_1, \hdots, u_{N_u}}^T$ for $u_n = n / (N_u + 1)$, over uniform directions $\Omega \in \VEC{\Omega} = \CBRAK{\Omega_1, \hdots, \Omega_{N_{\Omega}} }$ on the unit-sphere. Our interpolated density $\VEC{D}_t$ at time $t$ minimizes the least-squares spherical sliced Wasserstein residuals (LS-SSWR)  w.r.t. the quantiles following
\begin{equation} \label{EQ:WASS_SLICE_LS_SSW}
\displaystyle
\begin{split}
\VEC{D}_t & = \argmin{\VEC{D}} \,  \sum_{i=1}^{N_{\Omega}} \,  \sum_{j = 1}^{N_u} \ABS{ F_{\Theta}(\theta_{ij} \, | \, \MAT{R}_{\Omega_i} \VEC{D} ) - u_j}^2,  \\
\theta_{ij} & = (1 - t) \,  F_{\Theta}^{\minus 1} \PAREN{u_j \, | \,  \MAT{R}_{\Omega_i} \VEC{D}_0 } + t \, F_{\Theta}^{\minus 1} \PAREN{u_j \, | \,    \MAT{R}_{\Omega_i} \VEC{D}_1}, 
\end{split}
\end{equation}
where $F_{\Theta}(\theta_{ij} \, | \, \MAT{R}_{\Omega_i} \VEC{D} ) = \VEC{a}^T_{\theta_{ij}} \,  \MAT{R}_{\Omega_i} \VEC{D}$ is the product of vector $\VEC{a}^T_{\theta_{ij}}$ containing the marginal CDF terms in Eq. \eqref{EQ:MARGIN_CDF} given by
\begin{equation} \label{EQ:WASS_SLICE_WTS}
\displaystyle
\begin{split}
 \VEC{a}_{\theta_{ij}} & = \BRAK{a_0^0(\theta_{ij}), a_{\minus 1}^1 (\theta_{ij})  \hdots, a_{L_D}^{L_D} (\theta_{ij})}^T  \in \field{R}^{N_D \times 1}, \\
 a_l^0 (\theta_{ij})  & =   \sqrt{\frac{ \pi }{ 2l + 1 }}  \left \{
  \begin{array}{cc}
1 -  \cos\theta_{ij}, &  l = 0 \\
 P_{l \minus 1} \PAREN{\cos\theta_{ij} } -  P_{l \plus 1} \PAREN{\cos\theta_{ij} }, &  l > 0 
 \end{array}
\right .  ,
\end{split}
\raisetag{10ex}
\end{equation}
and $a_l^m  (\theta_{ij})  = 0, \, \forall m \neq 0$. Note that LS-SSWR expansions do not guarantee a non-negative density $f_{\VEC{D}_t}(\theta, \phi)$; increasing the resolutions of $\Omega$ and $u$ reduces but does not eliminate any negative regions over $(\theta, \phi)$. We therefore redress the formulation as follows:

Consider the non-negative SH expansion evaluations $\VEC{f} = \MAT{Y}_{\VEC{\theta} \VEC{\phi}} \, \VEC{D} \in \field{R}^{N_D \times 1}_{\geq 0}$ at uniform spherical coordinates $\VEC{\theta}, \VEC{\phi} \in \field{R}^{N_D}$, and invertible expansion matrix $\MAT{Y}_{\VEC{\theta} \VEC{\phi}} \in \field{C}^{N_D \times N_D}$.
We can recover $\VEC{f}$ by substituting $\MAT{D} =  \MAT{Y}^{\minus 1}_{\VEC{\theta}  \VEC{\phi}} \, \VEC{f}$ into Eq. \eqref{EQ:WASS_SLICE_LS_SSW}, and minimizing the following NNLS-SSWR problem w.r.t. the unknown density evaluations $\VEC{f}$:
\begin{equation} \label{EQ:WASS_SLICE_NNLS}
\displaystyle
\begin{split}
\VEC{f}_* & = \argmin{\VEC{f}} \,  \NORM{ \PAREN{\MAT{W} \MAT{Y}^{\minus 1}_{\VEC{\theta}  \VEC{\phi}} }  \VEC{f}  - \VEC{1} \otimes \VEC{u} }^2 \quad \textrm{s.t.} \quad 
\VEC{f} \geq \VEC{0},  \\
\MAT{W} & = \BRAK{\VEC{w}_{11}, \VEC{w}_{12}, \hdots, \VEC{w}_{N_{\Omega} N_u } }^T \in \field{R}^{N_{\Omega} N_u \times N_D}, \quad
\VEC{w}_{ij}  = \MAT{R}_{\Omega_i}^T \VEC{a}_{\theta_{ij}},
\end{split}
\raisetag{8.5ex}
\end{equation}
where $\otimes$ is the Kronecker product operator. As a result, the minimizer $\VEC{f}_*$ contains the supports of a SOMS density expansion $f_{\VEC{D}}(\theta, \phi)$ in Eq. \eqref{EQ:SH_DENSITY_NON_NEG}. 
We therefore solve for SOMS-SH coefficients that minimize the least-squares expansion error w.r.t. $\VEC{f}_*$ via the following SOS-SDP:
\begin{equation} \label{EQ:WASS_SLICE_SDP}
\displaystyle
\begin{split}
\bar{\VEC{Q}}_* & = \argmin{\bar{\MAT{Q}}}  \sum_{n=1}^{N_D} y_n^2 (\bar{\MAT{Q}})
 \qquad  \textrm{s.t.} \qquad 
 \bar{\MAT{Q}} \succeq 0, \\ 
 y_n (\bar{\MAT{Q}}) & =  \trace{ \bar{\MAT{Q}} \PAREN{  \bar{\VEC{Y}}_{\theta_n \phi_n} \bar{\VEC{Y}}_{\theta_n \phi_n}^T } } - f_{*n},	\quad  
 \bar{\MAT{Q}}_*  = \MAT{V} \MAT{\Sigma} \MAT{V}^T,
\end{split}
\raisetag{9.75ex}
\end{equation}
where the unknown matrix $\bar{\MAT{Q}} \in \field{R}^{N_C \times N_C}$ is constrained to be positive semidefinite, and the minimizer is post-normalized $\trace{\bar{\VEC{Q}}_*} = 1$ to have unity integral density. The SOMS-SH expansion coefficients are similarly recovered as in SOS-SDP Eq. \eqref{EQ:SH_DENSITY_LIKELIHOOD_OPT} via the weighted eigenvectors $\bar{\MAT{C}}_* = \MAT{V} \MAT{\Sigma}^{\frac{1}{2}}$ of $\bar{\MAT{Q}}$,  and sum of SH product operators $\bar{\MAT{D}}_* = \sum_{n=1}^{N_C} \bar{\VEC{C}}_{*n} \diamond \,  \bar{\VEC{C}}_{*n}$. 
Interpolated densities shown in Fig. \ref{FIG:DISP_INTERP} smoothly track mass displacement over time-steps from an initial dirac-delta like source to complex target distributions. We now apply our SSW interpolations to SRIR echo densities.

\section{Experiments}
\label{SEC:EXP}

In moving sound-source simulations, let us compute the RIRs of a common acoustic source placed at discrete points along a path trajectory.
For each source-position uniformly distributed along a line-path between coordinates $\BRAK{1, \minus 1, 1}$ and $\BRAK{1, 1, 0}$ in Fig. \ref{FIG:RIR_EXP}, we compute the SH-ISM RIRs \cite{luo2021FSRR} of image-source contributions to a receiver at coordinates $\BRAK{0.1, 0.3, 0.5}$ in a rectangular room of size $\BRAK{7, 8, 5}$ meters with uniform wall-reflection coefficients of $1/3$.
We expand each of the $n^{th}$ image-sources with incident angle $(\theta_n, \phi_n)$ via a weighted SOMS-SH projection kernel in Eqs. \eqref{EQ:SH_DENSITY_NON_NEG}, \eqref{EQ:SH_PROJ} upto max-order $L=8$; weights are proportional to the product of each image-source's inverse distance to the receiver and its acoustic wall-reflection attenuation. Integrating each SH-ISM RIR over its simulation duration ($150$ ms) yields the ground-truth ISM densities indexed by the normalized path time $t$ in Fig. \ref{FIG:RIR_EXP} (top row).

Next, we compare the ground-truth densities to the following interpolated densities conditioned on the end-points $t = 0$, $t = 1$: 
Linear $(1-t) f_{\VEC{D}_0}(\theta, \phi) + t f_{\VEC{D}_1}(\theta, \phi)$, product $f_{\VEC{D}_0}(\theta, \phi)^{1 \minus t} f_{\VEC{D}_1}(\theta, \phi)^{t}$, and Wasserstein Eqs. \eqref{EQ:WASS_SLICE_NNLS}, \eqref{EQ:WASS_SLICE_SDP} are post-normalized to have unity integrals.
The SSW distances of Eq. \eqref{EQ:WASS2_1D} for varying $t$ are shown in Fig. \ref{FIG:RIR_EXP} whereby the Wasserstein interpolation achieves the best generalization across the entire path. Lowering the max-order expansion $L \in \CBRAK{8, 4, 2}$ decreases the relative linear/Wasserstein distances at $t=0.5$ to $\CBRAK{1.51, 1.35, 1.18}$. 
Interpolated densities at $t=0.5$ illustrate qualitative differences; linear density fades between the direct source contributions, product density filters for mutual contributions, and Wasserstein density tracks the moving source and reflection contributions over the $2$ meter discretization.

\section{Conclusions}
\label{SEC:CONC}

We introduced SOMS-SH expansions for estimating spatial band-limited density functions of aggregate acoustic source and reflection directions. Maximum likelihood SDP optimizers were derived and tightened from quadratic to linear number of variables, and efficient inverse transform sampling methods exploited analytic marginal and condition CDF formulation of SH expansions. This accelerated computations of SSW distances, which made the least-squares Wasserstein interpolation tractable. Our NNLS-SDP formulation produced valid interpolated densities, and generalized moving SRIR SH-ISM densities over a line-path compared to conventional interpolants. As a result, acoustic source and reflection densities were inferred from coarser measurement grid discretizations in a room.

\pagebreak
\clearpage



\section{Compliance with Ethical Standards}
This is a numerical simulation study for which no ethical approval was required.

\bibliographystyle{IEEEbib}
\bibliography{strings,refs}

\end{document}